\documentclass[twocolumn,aps,superscriptaddress,prb]{revtex4-2}

\usepackage{bm}%
\usepackage{float}
\usepackage{svg}
\expandafter\ifx\csname package@font\endcsname\relax\else
 \expandafter\expandafter
 \expandafter\usepackage
 \expandafter\expandafter
 \expandafter{\csname package@font\endcsname}%
\fi

\usepackage{comment}
\usepackage{array}
\usepackage{amsmath,amssymb}
\usepackage{MnSymbol}
\usepackage{tikz-feynman,contour} 
\usetikzlibrary{shapes,arrows,positioning,automata,backgrounds,calc,er,patterns}
\usepackage{feynmf}
\tikzfeynmanset{compat=1.0.0} 
 
\usepackage{graphicx}
\usepackage{mathrsfs}
\usepackage{bm}
\usepackage{mathtools}
\usepackage{epstopdf}
\usepackage{hyperref}

\hypersetup{%
   pdfpagemode=None, 
   pdfstartpage=1,
   pdfmenubar=true,
   pdftoolbar=true,
   colorlinks = true,
   linkcolor=blue,
   citecolor=blue,
   urlcolor=blue,
   bookmarksopen=false
 }
 \usepackage{amsmath}
\usepackage{amsfonts}
\usepackage{mathtools}
\usepackage{graphicx}
\usepackage{fixme}
\usepackage{color}
\usepackage{cancel}
\usepackage{bm}

\usepackage{epstopdf}
\usepackage{float}
\usepackage{amsmath}
\usepackage{color}
\usepackage{comment}
\usepackage{dsfont}
\usepackage{braket}
\usepackage[normalem]{ulem}

\newcommand\identity{1\kern-0.25em\text{l}}

\usepackage{feynmf}
\usepackage{tikz-feynman}
\usetikzlibrary{shapes,arrows,positioning,automata,backgrounds,calc,er,patterns}
\tikzfeynmanset{compat=1.0.0}

\newcommand{\Up}{{\uparrow}}
\newcommand{\Dn}{{\downarrow}}

\newcommand\bwt         {\begin{widetext}}
\newcommand\ewt         {\end{widetext}}

\def\bi{{\bf i}}
\def\bj{{\bf j}}

\def\bk{{\bf k}}
\def\bp{{\bf p}}
\def\bq{{\bf q}}
\def\bk{{\bf k}}
\def\bK{{\bf K}}

\def\br{{\bf r}}

\definecolor{airforceblue}{rgb}{0.36, 0.54, 0.66}
\definecolor{amber}{rgb}{1.0, 0.75, 0.0}
\definecolor{applegreen}{rgb}{0.55, 0.71, 0.0}
\definecolor{alizarin}{rgb}{0.82, 0.1, 0.26}

\begin{document}

\title{Probing a quantum spin liquid with  equilibrium and nonequilibrium hole dynamics}
\author{J.\ H.\ Nyhegn}
\affiliation{Center for Complex Quantum Systems, Department of Physics and Astronomy, Aarhus University, Ny Munkegade, DK-8000 Aarhus C, Denmark}
\author{K.\ Knakkergaard Nielsen}
\affiliation{Max-Planck Institute for Quantum Optics, Hans-Kopfermann-Str. 1, D-85748 Garching, Germany}
\author{G.\ M.\ Bruun}
\affiliation{Center for Complex Quantum Systems, Department of Physics and Astronomy, Aarhus University, Ny Munkegade, DK-8000 Aarhus C, Denmark}

\begin{abstract}
The properties and experimental identification of quantum spin liquids (QSLs) remain an important topic with many fundamental questions. Here, we explore the dynamics of a single charge dopant (hole) in a $t$-$J_{1}$-$J_{2}$ model on a square lattice, which realizes a gapless $\mathbb{Z}_2$ QSL at half filling. Using a field theory approach based on the parton construction,
which includes an infinite number of scatterings between the low-energy quasiparticle excitations of the QSL via a self-consistent Born approximation, we calculate both the equilibrium and nonequilibrium properties of the hole for weak and strong interactions. Quasiparticle branches as well as stringlike excitations of the holon are identified, and we furthermore explore the time-dependent spreading of a hole throughout the QSL after it has been injected at a given lattice site. The final ballistic expansion speed is shown to exhibit a nonmonotonic behavior as a quantum phase transition between an antiferromagnetic and the QSL phase is crossed, which is caused by a qualitative change in the fundamental kinematics of the interactions between the hole and the surrounding spins. Our results demonstrate how charge dopants can be used as a quantum probe for QSLs and are directly relevant to optical lattice experiments with single site resolution. 
\end{abstract}

\maketitle

\section{Introduction} 
The fundamental nature of quantum spin liquids (QSLs) continues to attract great interest since these phases were first identified more than fifty years ago~\cite{anderson1973}. Quantum spin liquid phases such as resonating valence bond states are typically 
described as a fluid of strongly correlated spin singlets. The liquid or resonating character of the ground state means that it consists of a macroscopically large superposition of configurations with spin singlets covering the lattice, which can give rise to exotic properties such as long-range quantum entanglement, quasiparticle excitations with nontrivial statistics, and spin-charge separation~\cite{savary2017, broholm2020, kalmeyer1987, lee2006, knolle2019}. Thanks to sophisticated analytical and numerical methods,
 our understanding of QSLs keeps improving~\cite{baskaran1987, wen2002, poilblanc2012, hu2013, meng2010, yan2011}.
 There is also great progress on the experimental side, where a spin liquid phase has recently been realized 
using the new powerful quantum simulation platform based on Rydberg atoms in optical tweezer arrays~\cite{semeghini2021}.
There are furthermore promising prospects of realising spin liquids in optical lattice 
experiments~\cite{Yamamoto_2020,Yang2021,sun2023,Lebrat2024,Prichard2024}. Finally, the number of strongly 
correlated two-dimensional (2D) materials is growing including  atomically thin 
 transition metal dichalcogenide bilayers, which  realize highly tunable moir\'e lattice systems where spin liquids are predicted to emerge~\cite{Pan2020,Zang2021,MoralesDuran2022}.
\begin{figure}[hbt!]
	\begin{center}
	\hspace{-0.3cm}
	\includegraphics[width=0.41\textwidth]{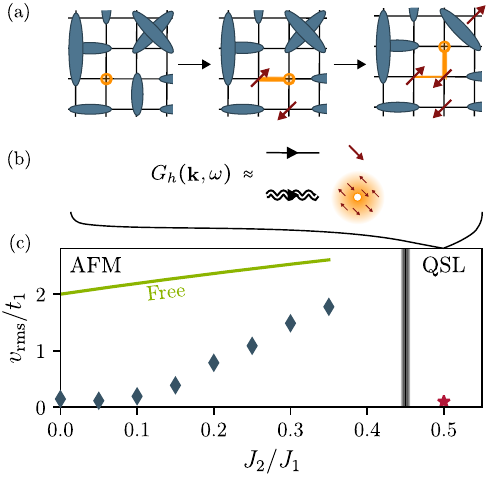}
	\end{center}
	\caption{(a) As a holon (orange circle) moves through a QSL, it distorts the spin singlets in its surroundings (blue ovals) leaving unpaired spins in its wake (arrows). These processes give rise to quasiparticle states as well as stringlike excitations of the holon.   (b) The hole propagator is approximated by the uncorrelated copropagation of a dressed holon and a bare spinon. (c) The velocity of a hole spreading ballistically as a quasiparticle after being injected at a given lattice site. The expansion velocity increases with $J_2/J_1$ for $J_2/J_1<0.46$ (black vertical line) where the ground state is an AFM, but remains slower than a free particle (green line) with $v_{\text{rms}}/t_{1}= 2\sqrt{1+2(t_{2}/t_{1})^{2}}$, corresponding to a hole in a fully ferromagnetic background. For  $J_2/J_1>0.46$, the ground state is a $\mathbb{Z}_2$ QSL and the expansion velocity is much smaller. This gives rise to a nonmonotonic behavior of the velocity with a maximum likely to be close to the quantum phase transition point $J_2/J_1=0.46$. 
}
	\label{Fig.1}
\end{figure}

A feature of QSLs is that they elude local magnetic order down to zero temperature \cite{savary2017}. This invalidates standard approaches such as spin-wave theory in handling these strongly interacting phases. Except for the Kitaev spin liquid offering an exact solution \cite{kitaev2006}, this means that numerical tools such as variational Monte Carlo (VMC) and matrix product states (MPSs) have played an important role in exploring QSLs~\cite{jiang2012, hu2013, yan2011, iqbal2016, zhu2015}. Moreover, a smoking gun observation of a QSL remains a great challenge. Exact diagonalization and recent MPS studies suggest that the dynamical response from inserting charge dopants can be used as probes for these phases \cite{laeuchli2004, kadow2022, kadow2024}. 
Understanding the dynamics of dopants in spin liquids is however challenging and has mostly been approached phenomenologically or at a mean-field level \cite{laughlin1995, beran1996, lee2006, tang2013, rantner2001}. In these studies, the focus has furthermore  primarily been on the interactions with the gauge fluctuations. 

These experimental advances make it timely to further explore the dynamics of a  dopant in a QSL. 
In this paper, we focus on a hole in a so-called $\mathbb{Z}_2$ QSL realized in a $t$-$J_{1}$-$J_{2}$ model at half filling. We develop a field theoretical approach for analyzing this, which is capable of describing the strong coupling regime by including an infinite number of scatterings between the hole and its surrounding spins via a self-consistent Born approximation (SCBA). Using this, we identify quasiparticles as well as  stringlike excitations of the holon (See Fig. \ref{Fig.1}(a)). The resulting spectral function is shown to be quite different from that of a hole in an AFM, which indicates that it can be used to detect the presence of a QSL. We furthermore explore the nonequilibrium spreading of a hole throughout the lattice ensuing its injection at a specific lattice site.  For long times, the hole is shown to expand ballistically as a quasiparticle whose speed depends nonmonotonically on the ratio $J_{2}/J_{1}$, with a maximum likely close to the quantum phase transition between the AFM and the QSL phases (See Fig. \ref{Fig.1}(c)). This nontrivial dependence provides another promising way to detect the QSL. 

The paper is organized as follows. In Sec.~\ref{sec.model}, we introduce the $t$-$J_{1}$-$J_{2}$ model. We then discuss the parton construction and  our field theoretical approach. The spectral properties of a holon are discussed in Sec.~\ref{sec.results}. In Sec.~\ref{sec.hole}, we explore the spectral properties of a physical hole as well as its nonequilibrium expansion dynamics. By comparing with the same properties of a hole in an AFM, we discuss in Sec.~\ref{sec.SCBA} two ways where the hole can be used as a quantum probe of a QSL. We end with conclusions and outlook in  Sec.~\ref{conclusions}.

\section{Model and field theory} \label{sec.model}
We study the behavior of a single dopant in a  $\mathbb{Z}_2$ QSL  in the paradigmatic $t$--$J$ model
with both nearest (NN) and next-nearest neighbor (NNN) interactions 
\begin{equation}
\hat{H}=\hat{H}_t+\hat{H}_J.
\label{Eq.TotalH}
\end{equation} 
Here
\begin{align}
	\hat{H}_t= & -t_{1}\sum_{\braket{\bi,\bj},\sigma}\tilde{c}^{\dagger}_{\bi,\sigma} \tilde{c}_{\bj,\sigma}  - t_{2}\sum_{\braket{\braket{ \bi,\bj }},\sigma}\tilde{c}^{\dagger}_{\bi,\sigma} \tilde{c}_{\bj,\sigma} + {\rm H.c.} 
	\label{Eq.Ht}
\end{align}
is the hopping Hamiltonian with NN and NNN  parameters $t_1, t_2$, and 
\begin{align}
\!\!\!\hat{H}_J = & \ J_{1}\sum_{\braket{\bi,\bj}} \left[ \hat{\mathbf S }_{\bi}\cdot \hat{\mathbf S }_{\bj} -\frac{1}{4}\hat{n}_{\bi}\hat{n}_{\bj} \right] + J_{2}\!\!\sum_{\braket{\braket{ \bi, \bj }}} \! \left[  \hat{\mathbf S }_{\bi}\cdot \hat{\mathbf S }_{\bi} -\frac{1}{4}\hat{n}_{\bi}\hat{n}_{\bj} \right]\!
	\label{Eq.HJ}
\end{align}
describes AFM Heisenberg type NN and NNN spin couplings $J_1, J_2 > 0$. The restricted fermionic operators $\tilde{c}^{\dagger}_{\bi,\sigma} = \hat{c}^{\dagger}_{\bi,\sigma}(1-\hat{n}_{\bi})$ ensure no double occupancy in terms of the onsite density operator $\hat{n}_{\bi}=\sum_\sigma\hat{c}^{\dagger}_{\bi,\sigma}\hat{c}_{\bi,\sigma}$. Here, $\hat c_{\bi\sigma}$ removes a fermion with spin $\sigma$ at site $\bi$ and $\hat{ {\bf S} }_{\bf i} = \frac{1}{2}\sum_{\sigma,\sigma'} \hat{ c }^\dagger_{{\bf i},\sigma}\boldsymbol{\sigma}_{\sigma\sigma'}\hat{ c }_{{\bf i},\sigma'}$. The  spin coupling in Eq. \eqref{Eq.HJ} naturally arises as an effective low-energy description of superexchange processes in the Fermi-Hubbard model for large onsite repulsion $U$ close to half filling, where $J_{1} = 4t_{1}^{2}/U$ and $J_{2} = 4t_{2}^{2}/U$. As this is relevant for strongly correlated 2D materials and experiments with atoms in optical lattices, we will in the following uphold these relations such that $J_{2}/J_{1}=(t_{2}/t_{1})^2$. To understand how the properties of the hole evolves with increasing interaction strength, we will, however, deviate from the limit $J_{i}/t_{i} \ll 1$.

At half filling, the Heisenberg model supports four distinct phases \cite{liu2022, wang2018, nomura2021, ferrari2020}. With no NNN interaction, the model exhibits antiferromagnetic (AFM) order. Turning on a nonzero $J_{2}$, we initiate a competition that frustrates the magnetic order, and by increasing $J_{2}/J_{1}$ the system  first enters a gapless $\mathbb{Z}_2$ QSL at $J_{2}/J_{1} \sim 0.46$, then a valence bond solid (VBS) phase at $J_{2}/J_{1} \sim 0.55$, and at last the magnetically ordered columnar phase at $J_{2}/J_{1} \sim 0.6$. To explore the QSL phase, one, thus, needs to be able to achieve NNN hoppings $|t_2|\simeq |t_1|/\sqrt{2}$. While NNN hoppings in standard optical lattice experiments are inherently much smaller than the NN hoppings, optical superlattices have been shown \cite{chalopin2024} to allow the desired control over $t_2 / t_1$. Focusing on the QSL phase, we now turn to the parton construction.

\subsection{Parton construction} \label{sec.parton}
\begin{figure}[t!]
	\begin{center}
	\includegraphics[width=0.48\textwidth]{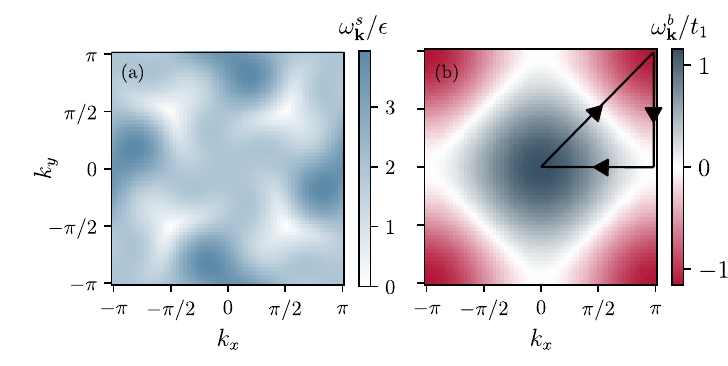}
	\end{center}
	\caption{(a) Spinon spectrum given by  Eq.~\eqref{Eq.spinon_spec} with $\Delta_{1}=1.1\epsilon$ and $\Delta_{2}=1.8\epsilon$. (b) Bare holon spectrum given by Eq.~\eqref{Eq.holondispersion}. }
	\label{fig.spinon.disp}
\end{figure}

Following Refs. \cite{song2021, brunkert2020, lee2006}, we decompose the fermions into partons as
\begin{align}
	\hat{ c }_{\bi,\sigma} =\hat{b}^{\dagger}_{\bi} \hat{f}_{\bi,\sigma},
	\label{eq.Composite}
\end{align}
where $\hat{b}_{\bi}$ removes a bosonic holon at site $\bi$ describing the charge degree of freedom, and $\hat{f}_{\bi,\sigma}$
removes a fermionic spinon describing the spin degree of freedom. 
Using the parton decomposition in Eq.~\eqref{Eq.HJ}, the mean-field theory most accurately describing the 
ground state  for $J_2 \simeq 0.5J_1$ at half filling with no holons is found to be  a gapless $\mathbb{Z}_2$ spin liquid \cite{wang2018, hu2013, ferrari2018, ferrari2020, yu2018}. From VMC results,  the optimal mean-field parton Hamiltonian was found to be 
\begin{align}
	\hat{H}_{J} =  \sum_{\bk, \sigma} \epsilon_{\bk} \hat{ f }^{\dagger}_{\bk,\sigma} \hat{ f }_{\bk,\sigma}  + \sum_{\bk} \left( \Delta_{\bk} \hat{f}^{\dagger}_{\bk,\uparrow} \hat{f}^{\dagger}_{-\bk,\downarrow} + \text{h.c.}\right), 
	\label{Eq.spinon} 
\end{align}
with the BCS-type pairing functions~\cite{wen2002, yu2018, hu2013}
\begin{align}
\epsilon_{\bk} &= \epsilon(\cos k_{x} + \cos k_{y}) \nonumber \\
\Delta_{\bk} &=\Delta_{1}(\cos k_{x}-\cos k_{y}) + \Delta_{2}\sin{(2k_{x})}\sin{(2k_{y})},\!\!
\label{eq.spinon_def}
\end{align}
where we have taken the lattice constant to unity.  Within the theory of the projective-symmetry group, this spin liquid is denoted Z2Azz13~\cite{wen2002}.
Instead of determining the parameters in Eq.~\eqref{eq.spinon_def}  via their self-consistent mean-field relations to the  Heisenberg parameters $J_{1}$ and $J_{2}$~\cite{wen2010},
we use unless otherwise stated the values $\Delta_{1}=1.1\epsilon$ and $\Delta_{2}=1.8\epsilon$,
found as the optimal parameters in the VMC study \cite{yu2018} for $J_{2}/J_{1}=1/2$. Note that Eq.~\eqref{Eq.spinon} is defined in an enlarged Hilbert space where unphysical doubly occupied sites are allowed keeping only the  average number of particles per site fixed at 
unity. These unphysical states can be projected out 
by using the Gutzwiller operator $\hat{\mathcal{P}} = \prod_{\bi}(\hat{n}_{\bi,\downarrow}-\hat{n}_{\bi,\uparrow})^{2}$ \cite{wen2010, iqbal2016, hu2013, ferrari2020}. It has however been shown that the the mean-field description given by 
Eq.~\eqref{Eq.spinon} without a subsequent Gutzwiller projection
accurately captures the low-energy excitations of the spin liquid state of the Heisenberg Hamiltonian $\hat H_J$ taking 
$\epsilon\simeq J_1/3$~\cite{ferrari2019}. Not having to perform a Gutzwiller projection dramatically simplifies the description, and we therefore use this mean-field approach in the following to describe the QSL. 
Diagonalizing Eq.~\eqref{Eq.spinon} then yields
\begin{align}
	\hat{H}_{J} =&   \sum_{\bk, \sigma} \omega^{s}_{\bk} \hat{\gamma}^{\dagger}_{\bk,\sigma} \hat{\gamma}_{\bk,\sigma}, 
	\label{Eq.spinon_diag}
\end{align}
with the spinon spectrum
\begin{align}
	\omega^{s}_{\bk} = \sqrt{\epsilon_{\bk}^{2}+\Delta_{\bk}^{2}},
	\label{Eq.spinon_spec}
\end{align}
shown in Fig. \ref{fig.spinon.disp}(a). The BCS-type ground state is defined by $\hat{\gamma}_{\bk,\sigma} \ket{\text{GS}} = 0$. 
Note that the Cooper pairs in this mean-field theory correspond to the spin singlets.

The hopping term $\hat H_t$ comes into play when we add a single dopant to the system.
Using the parton construction  yields
\begin{align}
	\hat{H}_{t} = -t_{1} \sum_{ \left\langle \bi, \bj \right\rangle } \hat{b}_{\bi} \hat{f}^{\dagger}_{\bi,\sigma} \hat{f}_{\bj,\sigma}  \hat{b}^{\dagger}_{\bj}  -t_{2} \sum_{\left\langle \left\langle \bi, \bj \right\rangle \right\rangle} \hat{b}_{\bi} \hat{f}^{\dagger}_{\bi,\sigma} \hat{f}_{\bj,\sigma}  \hat{b}^{\dagger}_{\bj},
	\label{Eq.Ht_parton}
\end{align}
where we have removed the projection operators since we only consider a single dopant and  allow multiple spinons on a given site~\cite{lee2006, song2021, kane1989}. Expressing this in terms of the spinon operators $\hat{\gamma}^{\dagger}_{\bk,\sigma}$ 
diagonalizing $\hat H_J$ giving Eq.~\eqref{Eq.spinon_diag}, we can write Eq.~\eqref{Eq.Ht_parton} as 
\begin{align}
	\hat{H}_{t} = &\sum_{ \bk }\omega^{b}_{\bk} \hat{b}^{\dagger}_{\bk} \hat{b}_{\bk}+  
	 \sum_{ \bk, \bk',\bq, \sigma } h_{\bk,\bk',\bq} \hat{b}^{\dagger}_{\bk-\bq} \hat{b}_{\bk} \hat{\gamma}^{\dagger}_{\bk', \sigma} \hat{\gamma}_{\bk'-\bq, \sigma} \nonumber \\ 
	+ & \sum_{ \bk, \bk',\bq }  g_{\bk,\bk',\bq} \left( \hat{b}^{\dagger}_{\bk-\bq} \hat{b}_{\bk} \hat{\gamma}^{\dagger}_{\bk', \uparrow}\hat{\gamma}^{\dagger}_{-\bk'+\bq, \downarrow} + \text{h.c.} \right).
	 \label{Eq.H_inter} 
\end{align}
Here 
\begin{equation}
	\omega^{b}_\bk = 2 \sum_{\bk'} \Gamma_{\bk+\bk'}  v_{\bk'}^{2} 
	\label{Eq.holondispersion}
\end{equation}
is the dispersion of a bare holon shown in Fig. \ref{fig.spinon.disp}(b), and 
\begin{align}
	g_{\bk,\bk',\bq} &= -\left[\Gamma_{\bk-\bk'}u_{\bk'} v_{\bk'-\bq} +  \Gamma_{\bk+\bk'-\bq}  v_{\bk'} u_{\bk'-\bq} \right] \nonumber  \\ 
	h_{\bk,\bk',\bq} &=\Gamma_{\bk-\bk'} u_{\bk'} u_{\bk'-\bq} -\Gamma_{\bk+\bk'-\bq} v_{\bk'} v_{\bk'-\bq}  
 \label{Eq.vertex}
\end{align}
are the vertex functions with 
\begin{align}
	\Gamma_\bk =& \frac{2}{N} ( t_{1}\left[ \cos{k_{x}} + \cos{k_{y}} \right] \nonumber \\ 
	+ &t_{2}\left[ \cos{(k_{x}+k_{y})} + \cos{(k_{y}-k_{y})} \right] )
\end{align}
coming from NN and NNN structure factors. The number of lattice sites is $N$, and 
the coherence factors are  $u_{\bk} = \sqrt{( 1 + \epsilon_{\bk}/\omega^{s}_{\bk}) / 2}$ and 
$v_{\bk} = \text{sgn}\Delta_{\bk}\sqrt{( 1 - \epsilon_{\bk}/\omega^{s}_{\bk}) / 2}$. 
There are two kinds of interaction vertices between the holon and the spinons in Eq.~\eqref{Eq.H_inter}:
$h_{\bk,\bk',\bq}$ describes the scattering of a holon on a spinon whereas 
$g_{\bk,\bk',\bq}$ describes 
the emission/absorption of two spinons by a holon. 

With Eqs.~\eqref{Eq.spinon_diag} and \eqref{Eq.H_inter}, we have arrived at our final Hamiltonian describing a 
dopant in a $\mathbb{Z}_2$ QSL.
Staying true to the Fermi-Hubbard relation $J_i = 4t_i^2/U$, we use $t_{2}/t_{1}= -\sqrt{J_2/J_1} = -1/\sqrt{2}$ throughout 
corresponding to the case of a single hole doping ($t_{2}/t_{1}=1/\sqrt{2}$ for a doublon \cite{jiang2021c}).
We also keep the ratio $\Delta_2/\Delta_1=1.8/1.1$ fixed and  $\Delta_{1}=1.1\epsilon$ unless otherwise stated, 
which leaves  $\epsilon/t_{1}$ 
as a free  parameter in our theory. Strong interaction  with the holon creating many spinon excitations 
corresponds to $\epsilon/t_{1}\ll 1$ whereas $\epsilon/t_{1}\gg 1$ corresponds to weak interaction. 
 
\subsection{Quantum field theory} \label{sec.QFT}
\begin{figure}[t!]
	\begin{center}
	\includegraphics[width=0.3\textwidth]{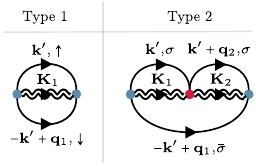}
	\end{center}
	\caption{Feynman diagrams included in the extended SCBA for the holon self-energy. Double wavy lines are the full holon propagator and  the single  lines are  spinon propagators. The blue and red dots are the $g$ and $h$ vertices in 
 Eq.~\eqref{Eq.vertex} respectively.}
	\label{Fig.diagrams}
\end{figure}

We will now describe our quantum field theory approach  to describe the motion of the hole in the QSL. 
In the parton construction, a first natural step is to analyze the motion of the holon 
quasiparticle~\cite{laughlin1995, rantner2001} as described with the retarded holon Green's function 
\begin{equation}
	G(\bk,\tau) = -i\theta(\tau)\langle \{ \hat{b}_{\bk}(\tau), \hat{b}^{\dagger}_{\bk}(0)\}\rangle,
	\label{Eq:Greens}
\end{equation}
where $\hat A(\tau)=\exp(i\hat H\tau) \hat A\exp(-i\hat H\tau)$ is the operator at time $\tau$ in the Heisenberg picture and 
$\{\hat A,\hat B\}=\hat A\hat B+\hat B\hat A$. In frequency/momentum space, a formal solution for the Green's function is 
\begin{align}
	G(\bk,\omega) = \frac{1}{\omega - \omega^{b}_{\bk} - \Sigma(\bk,\omega) + i0^{+}},
	\label{Eq:Greens}
\end{align}
where  $\Sigma(\bk,\omega)$ is the holon self-energy. In order to calculate the self-energy, one in general needs to resort to approximations. Here, we will apply the self-consistent Born  approximation (SCBA)~\cite{kane1989}, which has proven to be very accurate for describing 
a mobile hole in an AFM even for strong interactions when compared to Monte Carlo calculations and exact diagonalization~\cite{martinez1991, diamantis2021}. Remarkably, this holds even for describing nonequilibrium hole dynamics~\cite{nielsen2022}. 
In the present case with the two kinds of spinon-hole interaction vertices given by Eq.~\eqref{Eq.vertex}, 
the SCBA amounts to including the two types of diagrams for the holon self-energy shown in Fig. \ref{Fig.diagrams}.
These two diagrams give the exact solution of the self-energy to third order in the hopping amplitudes, 
and by using the self-consistent holon Green's function we include an infinite class of diagrams in order to describe 
the limit  $\epsilon/t_1\ll 1$ of strong interactions. 
The self-consistent diagrams in Fig.~\ref{Fig.diagrams} give the self-energy 
\begin{gather}
\Sigma(\bk,\omega) = \sum_{\bk',\bq_{1}}g_{\bk,\bk',\bq_{1}}^{2} G(\bK_{1},\omega-\omega^{s}_{\bk'}-\omega^{s}_{-\bk'+\bq_{1}})  \nonumber \\
-2\sum_{\bq_{2}} g_{\bk,\bk',\bq_1} G(\bK_{1},\omega-\omega^{s}_{\bk'}-\omega^{s}_{-\bk'+\bq_{1}})h_{\bk-\bq_{1},\bk'+\bq_{2},\bq_2}\nonumber\\
\times G(\bK_{2},\omega-\omega^{s}_{\bk'+\bq_{2}}-\omega^{s}_{-\bk'+\bq_{1}})  g_{\bk,\bk'+\bq_{2},\bq_{1}+\bq_{2}},\!\!
\label{Eq:self}
\end{gather}
with $\bK_{n} = \bk - \bq_{1} - ... - \bq_{n}$.
Thus, our field theoretical approach amounts to calculating the holon  Green's function  by solving Eqs.~\eqref{Eq:Greens} and \eqref{Eq:self} self-consistently, which  includes an infinite number of spinon excitations created by the holon.

\section{Holon properties} \label{sec.results}
We now describe the properties of a holon dynamics captured by its spectral function 
\begin{align}
	A(\bk,\omega) = -2\text{Im}\left[ G(\bk,\omega)\right],
\end{align}
which gives the energy spectrum of a holon injected with momentum $\bk$ into the spin liquid. 
As detailed in Appendix~\ref{app.OysterPlus}, it turns out that the second diagram in Fig.~\ref{Fig.diagrams}
has only small effects on the self-energy and since it is computationally quite demanding to evaluate,  it is omitted in the 
following. 

\begin{figure}[t!]
	\begin{center}
	\includegraphics[width=0.45\textwidth]{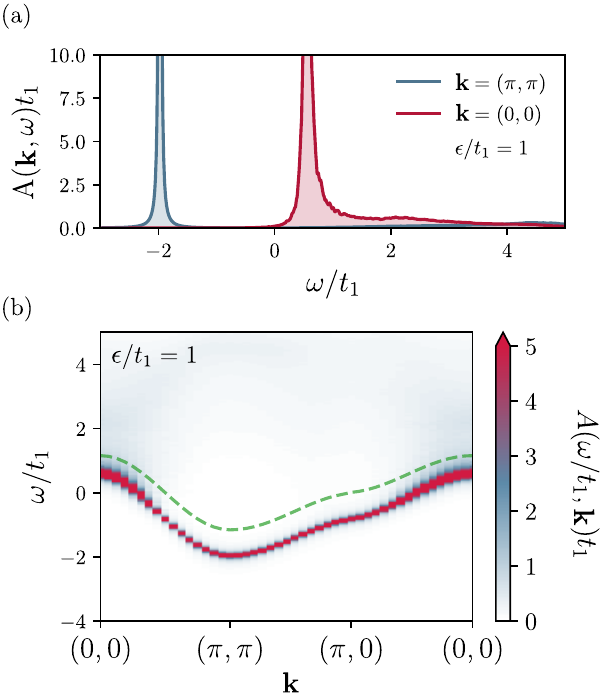}
	\end{center}
	\caption{(a) Spectral function of the holon for two different crystal momenta and the interaction strength  $\epsilon/t_{1}= 1$. (b) Spectral function for crystal momenta along the path in the Brillouin zone shown in Fig. \ref{fig.spinon.disp}(b). The green dashed line is the bare holon dispersion given by Eq.~\eqref{Eq.holondispersion}. }
	\label{Fig.spec1}
\end{figure}

Initiating the investigation in the weak coupling regime, we plot in Fig.~\ref{Fig.spec1}(a) the holon spectral function 
for $\epsilon/t_1=1$ and two different momenta obtained by numerically solving Eqs.~\eqref{Eq:Greens} and \eqref{Eq:self} 
self-consistently on a $32\times32$ lattice. We see that the spectral function exhibits  
 a clear quasiparticle peak for both momenta. Figure \ref{Fig.spec1}(b) shows that the quasiparticle is well defined for all  crystal momenta along the specified path in the Brillouin zone. The energy of this quasiparticle 
 is due to interactions with the spinons shifted by a constant relative to the bare holon dispersion given by the 
  green dashed line. We note that for all  results, 
the spectral function obeys the sum rule $\int d\omega A_h(\bp,\omega)=2\pi$ with less than 1\% deviation confirming 
the accuracy of our numerics.

\begin{figure}[t!]
	\begin{center}
	\includegraphics[width=0.48\textwidth]{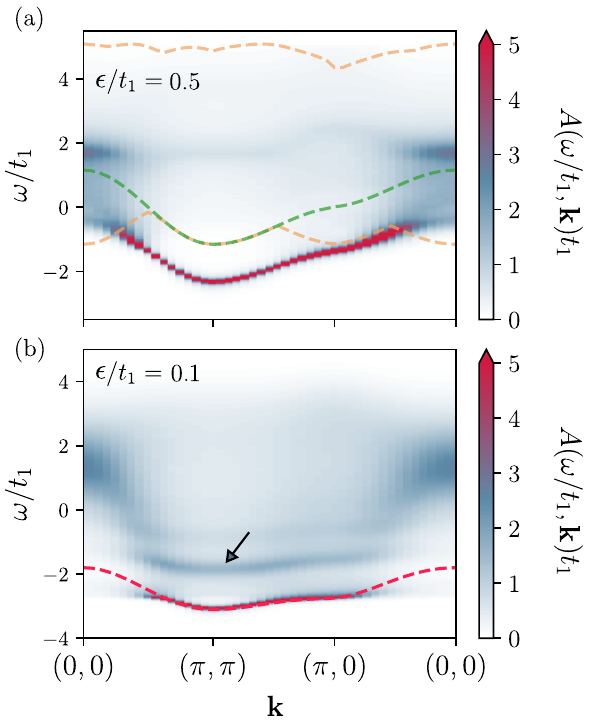}
	\end{center}
	\caption{Spectral function of the holon along the same path in the Brillouin zone as in Fig.~\ref{Fig.spec1}, but now for intermediate interaction strength $\epsilon/t_1=0.5$ (a) and strong  interaction strength $\epsilon/t_1=0.1$ (b). The dashed green line in (a) is again the bare holon dispersion given by Eq.~\eqref{Eq.holondispersion} whereas the orange lines indicate the boundaries of the spinon-holon many-body continuum. The red dashed line in (b) is a fit to the holon dispersion; see Appendix~\ref{app.QP.fit}.
 }
	\label{Fig.spec2}
\end{figure}

Figure \ref{Fig.spec2}(a) shows that for intermediate interactions $\epsilon/t_1=0.5$, the quasiparticle peak in the 
holon spectral function has lost significant spectral weight to a many-body continuum. Indeed, the quasiparticle 
is completely washed out for momenta around  $\bk = (0,0)$, which can be attributed to the emissions of spinon pairs
as described by the first Feynman diagram in Fig.~\ref{Fig.diagrams}. The on-shell energies of these 
emissions form a 
 continuum between $\text{min/max}\{ \omega^{b}_{\bk-\bq} + \omega^{s}_{\bk'} + \omega^{s}_{-\bk'+\bq}\}$, shown as 
 orange dashed lines in Fig.~\ref{Fig.spec2}(a). We clearly see that the quasiparticle is heavily damped when its energy is inside this 
 continuum. 
Increasing the interacting strength further to $\epsilon/t_1=0.1$, Fig.~\ref{Fig.spec2}(b) shows that the quasiparticle peak is 
damped even more due to decay into spinon modes as expected. We find that the  width of the quasiparticle band  is 
 approximately given by $3.9\epsilon/t_{1}$; i.e.,\ increasing the interaction strength decreases the bandwidth. 
 This reflects that increasing the 
 dressing of the holon by spinon modes results in a heavier quasiparticle. Such significant dressing cannot be captured by 
 mean-field approaches, which in  stark contrast predict a holon quasiparticle with a dispersion scaling as 
$\sim t_{1}$~\cite{laughlin1995, laughlin1997}. Our results are on the other hand consistent with MPS calculations for a chiral spin liquid predicting  similar features for the physical hole~\cite{kadow2022}. 
We will analyze the properties of a physical hole  in Sec.~\ref {sec.hole}.

\subsection{String excitations} \label{sec.string}
\begin{figure}[t!]
	\begin{center}
	\includegraphics[width=0.48\textwidth]{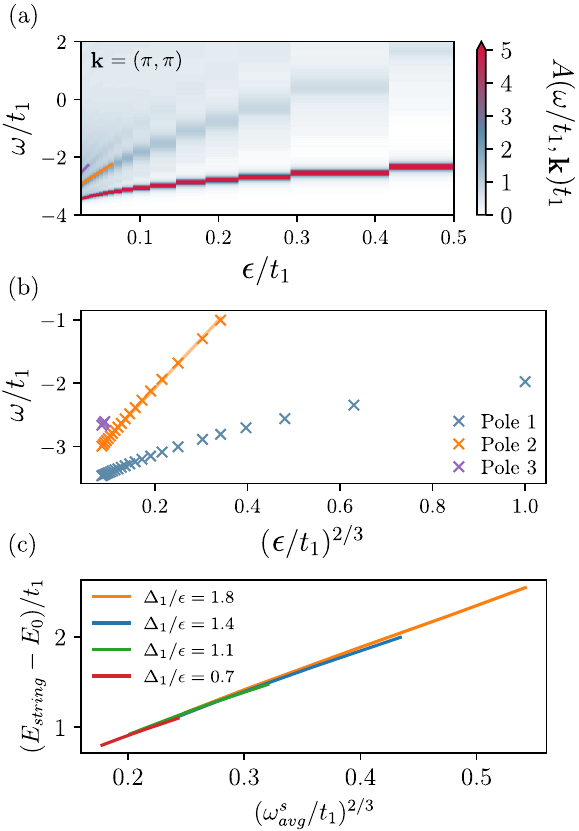}
	\end{center}
	\caption{(a) The holon spectral function for crystal momentum $(\pi,\pi)$ as a function of the interaction strength $\epsilon/t_1$. (b) Energy of the three lowest bands in (a) replotted as a function of $(\epsilon/t_{1})^{2/3}$. (c) Excitation energy $(E_{string}-E_0)/t_{1}$ of the lowest string band for momentum $(\pi,\pi)$ as a function of $(\omega^{s}_{avg}/t_{1})^{2/3}$.  $\omega^{s}_{avg}$ is the average energy of a spinon, and $E_0$ is the energy of excitation at $\epsilon/t_{1}=0$  }
	\label{Fig.string1}
\end{figure}

Intriguingly, one can also see several additional bands at higher energies in Fig.~\ref{Fig.spec2}(b). 
To investigate the origin of these bands, we plot in Fig.~\ref{Fig.string1}(a) the spectral function for momentum $\bp=(\pi,\pi)$ as a function of  $\epsilon/t_1$, clearly demonstrating that more excited states emerge with increasing interaction strength. Figure \ref{Fig.string1}(b) gives strong evidence that the energy of these excited states scale as $(\epsilon/t_{1})^{2/3}$. 
The ground state does not exhibit such a scaling coming from a continuum description, which could be due to its small spatial size making lattice effects important. 
For a hole in an AFM, such a scaling is interpreted as a smoking gun for the presence of so-called string excitations \cite{brinkman1970}. The basic mechanism is that the hole leaves a trace of frustrated spins in its wake as it moves through the AFM, which means that it experiences a linear potential. This gives rise to string states with the characteristic energy scaling in the continuum limit. Based on this, we conjecture that the bands in Fig. \ref{Fig.spec2}(b) also arise from string excitations of the holon. While there is no local magnetic order in a spin liquid, the holon still distorts its surroundings by creating pairs of spinon excitations as it moves though the lattice. This is  explicitly evident in Eq.~\eqref{Eq.H_inter} and Fig. \ref{Fig.diagrams}, and is illustrated in Fig. \ref{Fig.1}(a). The result is that the holon experiences a linear potential in real space leading to the excited string states visible in Fig.~\ref{Fig.spec2}(b). The resonating valence bond (RVB) background can repair the frustration created by the holon. The time-scale of this is given by $\epsilon$ 
whereas the time scale of the hopping is given by $t_{i}$. The string states must therefore be expected to emerge when 
$\epsilon/t_{i}\ll 1$, which indeed is what we observe. 
To support this hypothesis further, we plot in Fig.~\ref{Fig.string1}(c) the energies of the excited states as a function of the spinon excitation energy averaged over the Brillouin zone $\omega^{s}_{avg} = \sum_{\bk}\omega^{s}_{\bk}/N$. The different colors correspond to calculations with different ratios of $\Delta_{1}/\epsilon$. Figure \ref{Fig.string1}(c) shows that these different calculations collapse on a single line corresponding to the excitation energy scaling as $(\omega^{s}_{avg}/t_1)^{2/3}$.  This confirms that the low-energy states can be understood as string excitations coming from the holon breaking spin singlets into individual spinons so that it experiences a linear potential with a strength $\propto \omega^{s}_{avg}$. In Appendix~\ref{app.string}, we further investigate the nature of these string excitations and find that they are suppressed if one increases the average length of the spin singlets. 

\section{Properties of physical holes} \label{sec.hole}
We are now ready to explore the properties of the physical holes, which are the ones observable in experiments. To do this, 
we consider the hole Green's function 
\begin{gather}
	G_{h\sigma}(\bk,\tau) =-i\theta(\tau)\langle \{ \hat{c}_{\bk,\sigma}(\tau), \hat{c}^{\dagger}_{\bk,\sigma}(0)\}\rangle\nonumber\\
  =-i\theta(\tau)\langle  \hat{c}_{\bk,\sigma}(\tau) \hat{c}^{\dagger}_{\bk,\sigma}(0)\rangle=\theta(\tau)G_{h}^>(\bk,\tau).
  \label{GreensHole}
\end{gather}
The second identity follows from the fact that only the single hole injected is 
present in the lattice, and it is crucial  for enabling us to describe the nonequilibrium 
dynamics without having to use more cumbersome Keldysh techniques, as will be exploited in Sec.~\ref{RealspaceDyn}. 

Using the parton construction, the hole Green's function becomes a two-body Green's function   
\begin{align}
	G_{h}(\bk,\tau) = -\frac{i}{N}\!\sum_{\bq_{1},\bq_{2}}\!\left\langle \hat{f}^{\dagger}_{\bk+\bq_{2},\sigma}(\tau) \hat{b}_{\bq_{2}}(\tau) \hat{b}^{\dagger}_{\bq_{1}}  \hat{f}_{\bk+\bq_{1},\sigma} \right\rangle,\!
	\label{Eq:prop_hole}
\end{align}
 describing a spinon-holon "particle-hole" bubble. Here and in the following, we suppress the spin index on the Green's function since it does not depend on the direction of the spin
 due to  the SU$(2)$ spin symmetry of the Fermi-Hubbard model. 
Writing it in terms of the rotated spinons gives
\begin{align}
	G_{h}(\bk,\tau) = \ &\frac{-i}{N}\sum_{\bq_{1},\bq_{2}} v_{\bk+\bq_{1}}v_{\bk+\bq_{2}} \nonumber \\  
	&\left\langle \hat{\gamma}_{-\bk-\bq_{2},\downarrow}(\tau) \hat{b}_{\bq_{2}}(\tau) \hat{b}^{\dagger}_{\bq_{1}}  \hat{\gamma}^{\dagger}_{-\bp-\bq_{1}, \downarrow}  \right\rangle.
	\label{eq.prop_hole_rot}
\end{align}
Since $G_{h}(\bk,\tau)$ is now transformed into a two-body  Green's function due to the parton construction, it is in general very challenging to calculate. As depicted in Fig. \ref{Fig.1}(b), we resort to approximating it with the lowest-order diagram 
\begin{align}
	\includegraphics[width=0.43\textwidth]{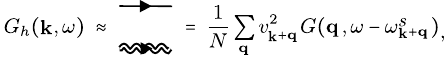}
	\label{eq.prop_hole_expre}
\end{align}
describing the uncorrelated motion of a bare spinon and a dressed holon. 
We find that this approximation obeys the frequency sum rule 
for the resulting hole spectral function within  $1\%$. 
Equation \eqref{eq.prop_hole_expre} is in fact in the spirit of the parton construction, where 
a physical particle (hole) is expected to split into a spin and charge part. This assumption may however 
break down in certain (small) regions of the Brillouin zone, which is a fundamental and interesting question to be 
examined in future work.

\subsection{Spectral properties} \label{sec.hole.spec}
\begin{figure}[t!]
	\begin{center}
	\includegraphics[width=0.45\textwidth]{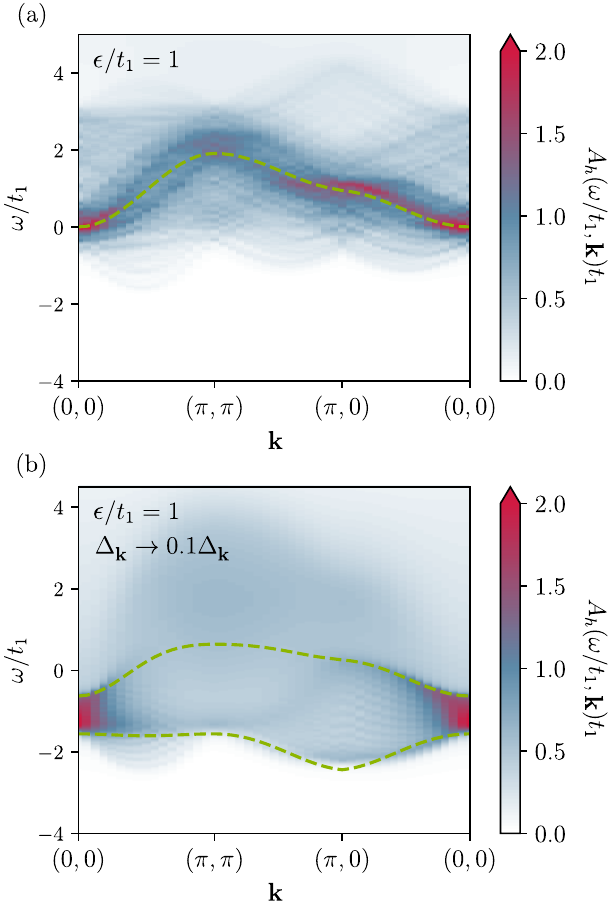}
	\end{center}
	\caption{ (a) Spectral function of a physical hole along the same path in the Brillouin zone as in Fig.~\ref{Fig.spec1}
 for interaction strength $\epsilon/t_1=1$ and $\Delta_1=1.1\epsilon$. 
The green dashed line is Eq.~\eqref{Holedispersion}.  In panel (b) the pairing field is taken to be ten times smaller with the 
top green dashed line showing Eq.~\eqref{Holedispersion} and the bottom green dashed line  
$ \omega = \omega^{\rm QP}_{\bk+(0,\pi)} + \omega^{s}_{(0,\pi)}$.}
	\label{Fig.Hole_1}
\end{figure}

To explore the spectral properties of the physical hole, measurable with ARPES \cite{kadow2022, kadow2024}, we plot in Fig.~\ref{Fig.Hole_1}(a) its spectral function
\begin{align}
	A_{h}(\bk,\omega) = -2\text{Im}\left[ G_{h}(\bk,\omega)\right]
	\label{eq.spec_hole}
\end{align}
calculated numerically from Eqs.~\eqref{eq.prop_hole_expre} and \eqref{eq.spec_hole} for the case of weak interactions   $\epsilon/t_1=1$. 
We clearly see a quasiparticle-like band, which however is significantly broadened compared to the corresponding holon quasiparticle band in Fig.~\ref{Fig.spec1}(b). It follows from Eq.~\eqref{eq.prop_hole_expre}
that the hole spectral function $A_{h}(\bk,\omega)$ is nonzero for energies satisfying 
 $\omega  \approx \omega^{\rm QP}_{\bq} + \omega^{s}_{\bk+\bq}$.
The dispersion of the hole band is in fact to a very good approximation given by 
\begin{align}
\omega^{h}_{\bk} = \omega^{\rm QP}_{\bk+\boldsymbol{\pi}} + \omega^{s}_{\boldsymbol{\pi}}
\label{Holedispersion}
\end{align}
with $\boldsymbol{\pi}=(\pi,\pi)$ as shown by the green dashed line in Fig.~\ref{Fig.Hole_1}(a). This can be understood from the fact that the spinon spectrum shown 
in Fig.~\ref{fig.spinon.disp}(a) has two Van Hove singularities at the center and corners of the Brillouin zone. 
While the coherence factor $v^2_{\bk+\bq}$ in Eq.~\eqref{eq.prop_hole_expre} vanishes at the origin suppressing its contribution to the hole spectral function, 
the Van Hove singularity at $\boldsymbol{\pi}$ gives rise to a large contribution to the hole spectral function,
which in turn gives a band with a dispersion given by Eq.~\eqref{Holedispersion}. 
 Similar features were reported in Ref. \cite{kadow2024}, where a hole in a Kitaev spin liquid was investigated using MPS methods.

In order to further explore how the spin environment affects the properties of the hole, we plot in Fig.~\ref{Fig.Hole_1}(b) its spectral function where we have used the same parameters as in Fig.~\ref{Fig.Hole_1}(a) except for a smaller pairing potential $\Delta_{\bk} \rightarrow 0.1\Delta_{\bk}$. 
Since the length of the singlets is determined by the BCS coherence length $\propto 1/\Delta$,
this makes them longer; see Appendix~\ref{app.string}. Figure \ref{Fig.Hole_1}(b) shows that the spectral function of the physical hole in turn becomes much broader with no clearly visible band except for small  momenta, reflecting the increased dressing of the underlying holon by spinons. The many-body continuum of the hole spectral function is approximately bounded from below by the energy $\omega^{\rm QP}_{\bk+(0,\pi)} + \omega^{s}_{(0,\pi)}$, 
which comes from a new Van Hove singularity in the spinon spectrum at $\bk=(0,\pi)$. These results show that the hole spectrum is broadened with less visible energy bands for spin liquids with longer singlets and a softer spinon spectrum. 

\begin{figure}[t!]
	\begin{center}
	\includegraphics[width=0.45\textwidth]{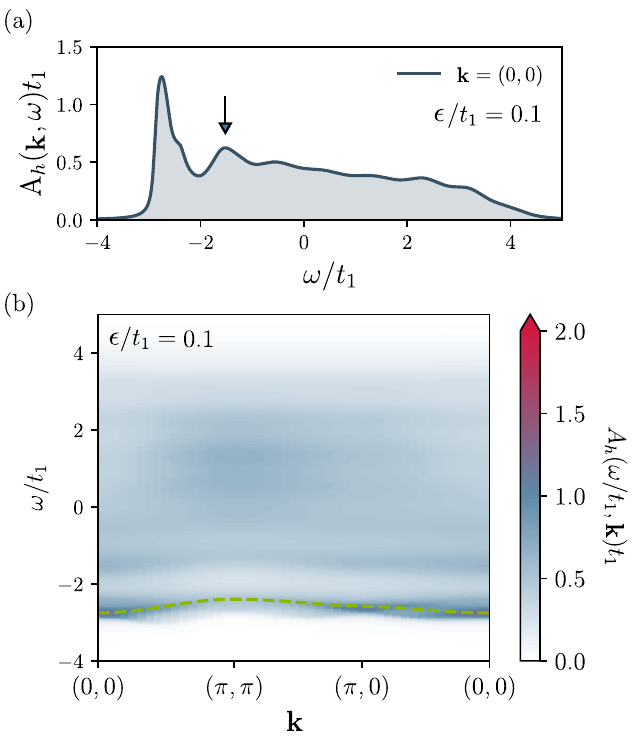} 
	\end{center}
	\caption{(a) Spectral function of a physical hole for momentum  $\bp= {\bf 0}$ and strong coupling 
 $\epsilon/t_{1}=0.1$. The arrow indicates a string excitation. In (b),
the spectral function is plotted  along the same path in the Brillouin zone as in Fig.~\ref{Fig.spec1}. The green dashed line  
shows Eq.~\eqref{Holedispersion}.}
	\label{Fig.Hole_3}
\end{figure}

In Sec.~\ref{sec.string}, we saw that the holon spectrum exhibits string excitations and we now investigate  whether they are present also for the physical hole. 
Figure \ref{Fig.Hole_3} shows the hole spectral function for strong interactions $\epsilon/t_1=0.1$ again taking $\Delta_1/\epsilon=1.8$ where the spectral function of the holon in Fig.~\ref{Fig.spec2}(b) exhibits clear string excitation bands. While the 
spectrum as expected is quite broad, there is still a band visible with an energy given by Eq.~\eqref{Holedispersion} shown by the green dashed line. 
Importantly, we also see distinct bands at higher energies and as analyzed further in  Appendix~\ref{app.hole.string}, they are a direct consequence of the string excitations of the holon. Hence, we conclude the physical hole spectral response can exhibit sharp 
lines, which arise due to the string like excitations of the holon. Such lines
have previously been interpreted as spinon-holon bound states~\cite{laughlin1995, beran1996, laughlin1997}.

\subsection{Real space nonequilibrium dynamics}\label{RealspaceDyn}
In a recent ground-breaking optical lattice experiment, the nonequilibrium motion of a hole after its release from a given lattice site in an 
AFM was explored in real space with single site resolution~\cite{ji2021a}. In light of this experiment and the prospects of realising spin liquids in optical lattices~\cite{Yamamoto_2020,Yang2021,sun2023,Lebrat2024,Prichard2024}, 
we therefore now turn to investigating the real-space dynamics following a sudden insertion of a hole.

Concretely, we calculate the hole Green's function $G_h^{>}(\br,\tau) = -i\langle \hat{c}_{\br,\sigma}(\tau)\hat{c}^{\dagger}_{{\mathbf 0},\sigma} \rangle$ in real space and time, 
which can be obtained from $G_{h}^>(\bk,\omega)$ by Fourier transforming. 
As discussed below Eq.~\eqref{GreensHole}, this gives us access to the  nonequilibrium dynamics described by the overlap of taking out a bare hole at site $\br$ and time 
$\tau>0$ after it was inserted at the origin at time $\tau=0$. 
Note that in order to describe  all dynamics, one needs the full time-dependent many-body wave function which is a much harder problem~\cite{nielsen2022}. 
However, the bare hole dynamics described by $G_h^{>}(\br,\tau)$ is sufficient to explore the long-time dynamics as has been 
successfully exploited both for magnetic polarons in AFMs~\cite{nyhegn2022,nielsen2022} and for polarons in atomic gases~\cite{skou2021a, nielsen2019}.

\begin{figure}[t!]
	\begin{center}
	\includegraphics[width=0.45\textwidth]{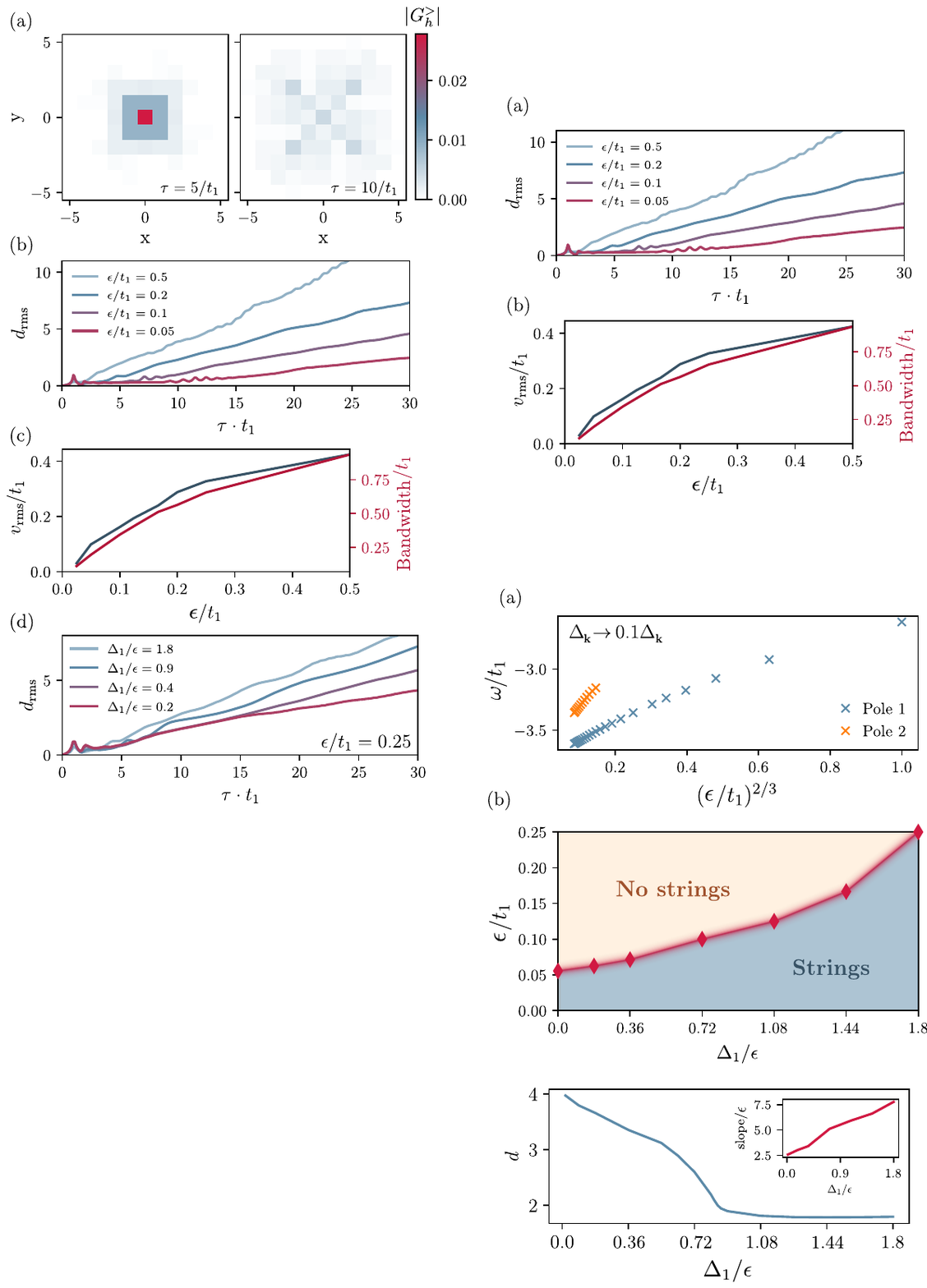} 
	\end{center}
	\caption{(a) The expansion of the hole throughout the lattice at times $\tau = 5/t_1$ (left) and $\tau = 10/t_1$ (right) after it was injected at the origin, as parametrized by $|G_h^{>}|$. The interaction strength is $\epsilon/t_{1}=0.25$. (b) The rms distance from its initial position  of an expanding hole  defined by Eq.~\eqref{Eq:rms} as a function of time for different interaction strengths. (c) The associated long-time expansion speed as a function of interaction strength (blue line). The red  line gives the bandwidth of the lowest hole quasiparticle band. (d) The rms distance as a function of time for the interaction strength $\epsilon/t_{1}=0.25$ and different pairing strengths $\Delta_1/\epsilon$ with $\Delta_{1}/\Delta_{2}$ constant.}
	\label{Fig.rms}
\end{figure}

In Fig.~\ref{Fig.rms}(a), we show an example of such calculations. We plot  
the the overlap $|G_h^{>}|$ in real space for two different times after the hole has been injected at the origin at time $\tau=0$ for the interaction strength $\epsilon/t_{1}=0.25$. This clearly shows how the hole spreads out through the lattice in a dynamical fashion. Note that the accumulated probability obtained by integrating the signal over the entire lattice is much larger in the left panel in Fig. \ref{Fig.rms}(a) than in the right. This reflects that the hole becomes increasingly dressed with spinons as it moves through the lattice, so that the total probability of observing a \emph{bare} hole decreases.
The slight symmetry breaking barely visible in the lower panel is an artifact of the gauge choice in the mean-field description of the QSL. 

In Fig.~\ref{Fig.rms}(b), we plot the gauge-invariant rms distance of the hole from its initial position at the origin 
defined as 
\begin{align}
	d_{\text{rms}}(\tau) = \sqrt{\frac{\sum_{{\bf r}} r^{2} |G^{>}_{h}({\bf r},\tau)  |^{2}  }{\sum_{{\bf r}} | G^{>}_{h}({\bf r},\tau)  |^{2} }}.
	\label{Eq:rms}
\end{align}
The denominator takes care of the decreasing probability to observe a bare hole with time. Figure \ref{Fig.rms}(b) shows that after some initial nontrivial dynamics, the hole expands ballistically  with a speed that decreases with increasing interaction strength. Physically, the reason is that the hole becomes increasingly dressed by spinons making it heavier. Indeed, similar results were found regarding the expansion of a hole in an AFM both 
theoretically~\cite{bohrdt2020,nyhegn2022,nielsen2022} and experimentally~\cite{ji2021a}.

Figure \ref{Fig.rms}(c) plots the final velocity, $v_{\mathrm{rms}} = \partial_\tau d_\text{rms}$, of this ballistic expansion as a function of the interaction strength, which clearly shows this decrease. We also plot in Fig.~\ref{Fig.rms}(c) the bandwidth of the lowest  band of the hole spectral function obtained from plots 
like those in Sec.~\ref{sec.hole.spec}. The agreement between these two quantities in Fig.~\ref{Fig.rms}(c) 
demonstrates that for long times, the hole expands as a quasiparticle with a dispersion determined by the lowest band in its spectral function. As this band becomes flatter with increasing interaction strength reflecting to dressing of the bare holon with spinons, it slows down. 

Again, deviating from the optimal VMC parameters by decreasing $\Delta_{1}/\epsilon$ while keeping $\Delta_{1}/\Delta_{2}$ constant, we see in Fig. \ref{Fig.rms}(d) that a spin environment with longer singlets slows down the expansion of the hole. This is due to longer singlets leading to more low-lying spinon excitations, which increase the dressing of the holon. It illustrates how the expansion of the hole can probe the spin environment in a QSL phase.
\section{The hole as a quantum probe}\label{sec.SCBA}
A central problem regarding QSLs is how to detect them. Armed with our field theoretical formalism capable of 
describing both equilibrium and nonequilibrium dynamics,  we will in this 
 section suggest two ways to use a hole as a quantum probe for spin liquids. 

In order to do this, we compare the spectral as well as expansion properties of a hole in a QSL with those  in an AFM, which is the ground state of the Heisenberg model $\hat H_J$ for small $J_2$~\cite{liu2022, wang2018, ferrari2018}. 
To calculate the properties of a hole in an AFM, we use linear spin wave theory combined with the SCBA, which has been 
shown to be remarkably accurate both for the equilibrium and nonequilibrium properties~\cite{kane1989, schmitt-rink1988,martinez1991, liu1991, marsiglio1991, liu1992, chernyshev1999, diamantis2021, nielsen2021, nielsen2022, nyhegn2022, nyhegn2023}. 
We refer the reader to Refs.~\cite{nielsen2021, nielsen2022, nyhegn2022, nyhegn2023} and Appendix~\ref{app.linear}
for the technical details of this method. 

\begin{figure}[t!]
	\begin{center}
	\includegraphics[width=0.45\textwidth]{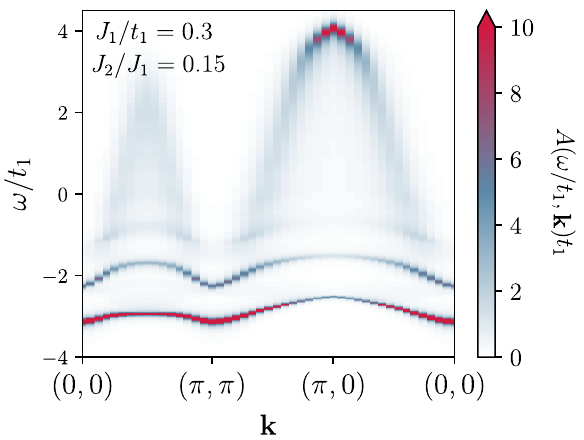} 
	\end{center}
	\caption{
 Spectral function of a hole along the same path in the Brillouin zone as in Fig.~\ref{Fig.spec1}
 for $J_{1}/t_{1}=0.3$, $J_{2}/J_{1}=0.15$, and $t_{2}/t_{1}=-\sqrt{J_{2}/J_{1}}$, where the system is in an 
AFM phase.}
	\label{Fig.spec.AFM}
\end{figure}

Let us first compare the spectral properties of a hole in an AFM  and in a QSL.
In Fig. \ref{Fig.spec.AFM}, we show the spectral function of the hole in an AFM  for 
$J_{2}/J_{1}=0.15$,  $t_{2}/t_{1}=-\sqrt{J_{2}/J_{1}}$, and $J_{1}/t_{1}=0.3$. 
Using $J_{1} \simeq 3\epsilon $ for the QSL~\cite{ferrari2019}, the latter ratio corresponds to $\epsilon/t_{1}\sim 0.1$. 
It is therefore natural to compare Fig. \ref{Fig.spec.AFM} with Fig. \ref{Fig.Hole_3}(b), which shows 
the spectral function of a hole in a QSL with $\epsilon/t_{1} = 0.1$ and $J_2/J_1=1/2$.
We see that these two spectral functions are qualitatively different. Whereas the spectral function in the 
AFM phase has several sharp quasiparticle bands corresponding to a ground state quasiparticle and string excitations, which 
take up significant spectral weight, the spectral function 
in the QSL phase has significantly broader bands with a many-body continuum taking up much more 
weight. From this we conclude that the spectral function of a hole provides a promising way to detect a spin 
liquid. Similar results were reported in Ref.~\cite{kadow2022}, where the spectral function for a hole in a
chiral spin liquid was calculated using MPS methods. 

Next, we compare the expansion dynamics of a hole in an AFM with that in a spin liquid, which can 
be measured in cold atom experiments. Our main result is shown in  Fig.\ \ref{Fig.1}, which plots the long-time ballistic 
expansion speed $v_\text{rms}$ as a function of the ratio $J_2/J_1$, where we have assumed that the 
spins form an AFM for $J_2/J_1<0.46$ indicated by the vertical black line, and a spin liquid for 
$J_2/J_1>0.46$~\cite{yu2018, wang2018, ferrari2018, nomura2021, liu2022}. We see that the expansion speed of the hole increases with 
$J_{2}/J_{1}$ in the AFM phase. This may seem surprising at first since increasing $J_{2}/J_{1}$ leads to a softer spin-wave spectrum
due to increased magnetic frustration, from which 
it would be tempting to conclude that the hole becomes more dressed and therefore slower in analogy with what happens in
the QSL. The reason this does not occur is that a nonzero value of $t_2 = -\sqrt{J_2/J_1} t_1$ 
means that the hole can hop to the next-nearest neighbors without distorting the AFM background. This gives rise to a 
bare hole dispersion, which starts to dominate the dynamics as the AFM order decreases with $J_2/J_1$ 
eventually leading to an increase in ballistic expansion of a 
bare hole. At a fundamental level, the hole interacts with the AFM background by emitting or absorbing a \emph{single}
spin wave~\cite{kane1989},
and these scattering processes become off-resonant when the characteristic energy of the bare hole dispersion is much larger than 
the spin wave energies, from which it follows that hole becomes less dressed. This is also indicated by the fact that the hole initially experiences a slight slowdown for increasing $J_{2}/J_{1}$. However, when the energy bandwidth of the bare hole exceeds the bandwidth of the spinwave spectrum, the dressing becomes inefficient and the hole speeds up.

The situation is qualitatively different in the QSL. Here, a basic interaction vertex corresponds to the emission or absorption of \emph{two} spinons as can be seen in Eq.~\eqref{Eq.H_inter} and Fig.~\eqref{Fig.diagrams}. This makes resonant spinon-holon scattering possible even when the characteristic energy of the spinon spectrum is 
much smaller than that of the bare holon. It follows that the holon and therefore the hole is always heavily dressed by spinons for strong interactions, even when the characteristic  energy of the bare holon dispersion
is large, which slows it down. While we unfortunately cannot 
 calculate the hole velocity for more values of $J_2/J_1$ inside the QSL, since we are only 
aware of numerical results providing a mapping to the mean-field description  for $J_2/J_1=1/2$~\cite{yu2018},
our results do allow us to predict that the ballistic expansion velocity of a hole depends \emph{nonmonotonically} 
on the  ratio $J_2/J_1$: In the AFM phase it increases whereas it becomes smaller again in the spin liquid phases as
can be seen in Fig.\ \ref{Fig.1}. The maximum velocity of the hole is likely to occur 
close to the quantum phase transition between the AFM and the QSL, and  
since this result is due to the different kinematics in the hole-medium scattering in the 
two phases, we believe it is robust and not an artifact of any approximations. Hence, we conclude that observing the  
expansion dynamics of a hole is a promising way to detect the transition between and AFM and a QSL.

\section{Conclusions and outlook}\label{conclusions}
Inspired by the remarkable advances with experimental platforms based on optical lattices and 2D quantum materials, we explored the equilibrium and nonequilibrium properties of a hole in a QSL. Using a field theory approach, we identified the presence of quasiparticle stringlike states of the hole, originating from breaking up singlets into their constituent spinons. The hole was furthermore shown to expand ballistically as a quasiparticle throughout the QSL for long times after its injection at a given lattice site, with a velocity exhibiting a characteristic nonmonotonic behavior as a quantum phase transition between the AFM and the QSL phases is crossed. As such, our results suggests promising new ways to use the hole as a quantum probe for QSLs. 

In addition to being relevant for new experiments exploring atoms in optical lattices and 2D quantum materials, our paper opens up several new research directions. This includes examining the possible presence of spinon-holon bound states, which in our field theory approach should show up as poles in the spinon-holon scattering matrix. It would be interesting to connect such calculations with earlier results stating that bound spinon-holon states are caused  by an attractive interaction mediated by gauge fluctuations~\cite{lee2006}. Another intriguing problem concerns the role of gauge fluctuations on the hole properties. While such gauge fluctuations likely are not important for the present $\mathbb{Z}_2$ QSL where they are gapped \cite{wen2002} and the low-energy spectrum of the mean-field description is similar to that after projection \cite{ferrari2019}, it would be interesting to compare with calculations performed using VMC \cite{charlebois2020} and MPS methods \cite{kadow2022,kadow2024}. Gauge fluctuations are moreover crucial for describing other QSLs such as those realized in triangular lattices~\cite{wen2002}. Finally, it would be very interesting to explore the binding of two holes in a QSL, which could lead to new and exotic superconductors/superfluids for a nonzero hole concentration.

\begin{acknowledgments}
This work has been supported by the Danish National Research Foundation through the Center of Excellence “CCQ” (Grant Agreement No. DNRF156), as well as by the Carlsberg Foundation through a Carlsberg Internationalisation Fellowship, Grant No. CF21\_0410. J.H.N. would like to thank Leon Balents and Federico Becca for useful comments and discussions.
\end{acknowledgments}

\clearpage


\appendix

\section{Extended self-consistent Born approximation}  \label{app.OysterPlus}
\begin{figure}[t!]
	\begin{center}
	\includegraphics[width=0.48\textwidth]{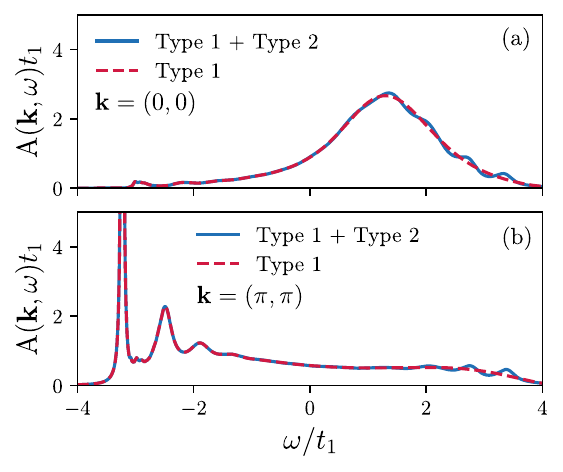}
	\end{center}
	\caption{Spectral function of the holon calculated using both types of diagrams from Fig. \ref{Fig.diagrams} (full line) and only type 1 (dashed line). These calculations are performed with  $\epsilon/t_{1} = 1/20$ for $\bk = (\pi,\pi)$ in panel (a), $\bk=(0,0)$ in panel (b), and a lattice size of 8x8. Except for small discrepancies at the top of the spectrum, we find that the overall behavior does not change by including the type 2 diagrams. This demonstrates the general tendency.}
	\label{fig.oyster.plus}
\end{figure}

In this appendix, we investigate what happens if we include the type 2 diagrams from Fig. \ref{Fig.diagrams} when we calculate the Green's function of the holon, Eq. \eqref{Eq:Greens}. Looking at types 1 and 2, we see that type 2 is a higher order diagram, with the order referring to the number of vertices. When including these diagrams, we, therefore, ensure that we keep the same order of diagrams. We do this by first including both types of diagrams in the iterative calculations, and then after 2/3 of the iterations we only include type 1. By including the type 2 diagrams, we cannot go much beyond a lattice size of 8x8, but in Fig. \ref{fig.oyster.plus} we see that including both types compared to only type 1 does not change the overall behavior of the spectral function. These two results represent the general tendencies.
For the calculation in the main text we hence only include the type 1 diagrams in order to perform the calculations for larger system sizes.

\section{String excitations}  \label{app.string}

\begin{figure}[t!]
	\begin{center}
	\hspace{-0.5cm}
	\includegraphics[width=0.38\textwidth]{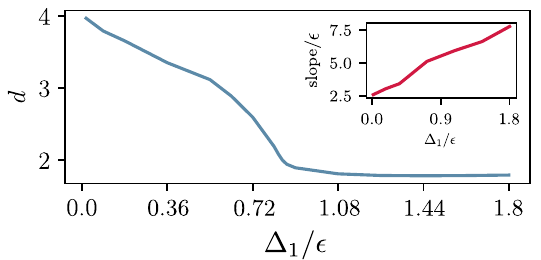}
	\end{center}
	\caption{Cooper pair length, Eq. \eqref{eq.avg_len}, for changing pairing potentials. As $\Delta_{\bk}/\epsilon$ is increased, the average length of the Cooper pairs is lowered and flattens out around $\Delta_{1}/\epsilon=1$. In the inset, we find the strength of string potential plotted as a function of $\Delta_{1}/\epsilon$. }
	\label{Fig.length}
\end{figure}

\begin{figure}[t!]
	\begin{center}
	\includegraphics[width=0.38\textwidth]{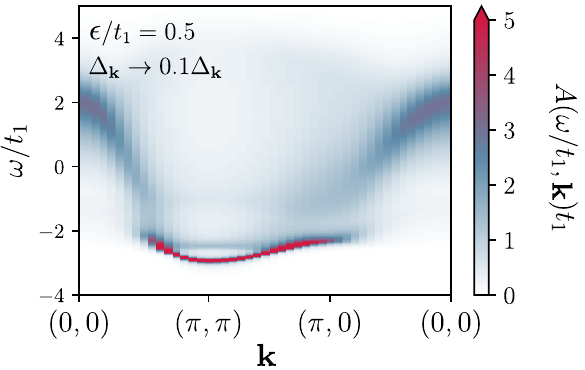}
	\end{center}
	\caption{Holon spectral function with $\epsilon/t_{1}= 0.5$, $t_{2}/t_{1} = -\sqrt{0.5}$, and $\Delta_{\bk} \rightarrow 0.1\Delta_{\bk}$ for a system size of 32x32. Comparing with Fig. \ref{Fig.spec2} we see that the QP bandwidth has decreased with a factor of $3$ and that fewer crystal momenta state has a well defined peak.}
	\label{Fig.spec_Delta}
\end{figure}

\begin{figure}[t!]
	\begin{center}
	\hspace{-0.5cm}
	\includegraphics[width=0.45\textwidth]{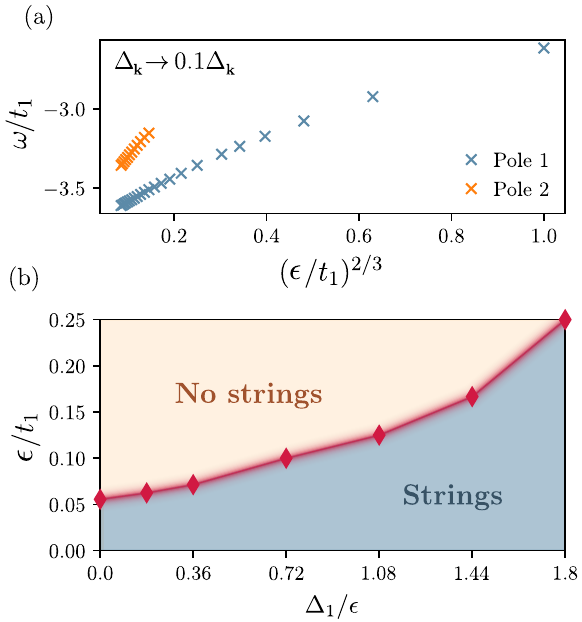}
	\end{center}
	\caption{Illustrating the importance of the pairing potential, we plot in panel (a) the poles of the Green's function as a function of the interaction strength with $\Delta_{\bk} \rightarrow 0.1\Delta_{\bk}$. The excited states do not show before $\epsilon/t_{1} \sim 1/16$ which is at a three times larger interaction strength compared to the results presented in Fig. \ref{Fig.string1}(b). Panel (b) summarizes the behavior by showing the interaction strength needed for the string excitations to show when the pairing potentials are changed. The red line illustrates when the string excitations show, and is the boundary between the blue region where no string excitations are present and the beige-colored region with string excitations present. }
	\label{Fig.delta}
\end{figure}

In this appendix, we further investigate the nature of the string excitations. We do this by tuning the pairing potentials $\Delta_{1}/\epsilon$ while keeping $\Delta_{1}/\Delta_{2}$ constant. In doing this the average excitation energy of the spinons, $\omega^{s}_{avg}$ decreases, and we retrieve a softer spinon spectrum. This results in a background with longer Cooper pairs and hence longer singlets, which is seen in Fig. \ref{Fig.length} by plotting the average length of the Cooper pairs,
\begin{align}
	d = \sqrt{\frac{\sum_{\mathbf{d}}|\Delta_{\mathbf{d}}|^{2}|\mathbf{d}|^{2}}{\sum_{\mathbf{d}}|\Delta_{\mathbf{d}}|^{2}}}, \ \Delta_{\mathbf{d}} = \frac{1}{N}\sum_{\br}\left\langle \hat{f}_{\br,\uparrow}\hat{f}_{\br+\mathbf{d},\downarrow} \right\rangle,
	\label{eq.avg_len}
\end{align}
as a function of $\Delta_{1}/\epsilon$. 
In an environment with more low-lying excitations, we also find a much heavier quasiparticle; see Fig. \ref{Fig.spec_Delta}. With the average cost of exciting the spinons decreasing, we would then expect the linear potential, given by the slope in Fig. \ref{Fig.string1}, to decrease. This slope is plotted in the inset in Fig. \ref{Fig.length}, and we see just that.
Another feature found is that, by decreasing the pairing potential, we need a larger interaction strength before the string excitations start to appear. In Fig. \ref{Fig.delta} we show results for $\Delta_{\bk} \rightarrow 0.1\Delta_{\bk} $ similar to those presented in Fig. \ref{Fig.string1}(b). 
Here, we see that the first excited states at an interaction strength below $\epsilon/t_{1} \sim 1/16$, and the ones corresponding to the third pole in Fig. \ref{Fig.string1}(a) do not show for the $\epsilon/t_{1}$ values shown. This behavior is summarized in Fig. \ref{Fig.delta}(b), indicating for which parameters string excitations occur. This, therefore, means that the tightly bound Cooper pairs promote the formation of these states. We believe the reason for this is the much heavier holon quasiparticle which hinders the formation of the strings.

\section{Linear spin wave theory}  \label{app.linear}
In this appendix, we will use the linear spin wave theory (LSWT) on the $t$-$J_{1}$-$J_{2}$ given by Eq. \eqref{Eq.Ht} and \eqref{Eq.HJ}. Considering a single dopant, we will proceed as done previously \cite{kane1989,martinez1991,liu1991,schmitt-rink1988,nielsen2021,nielsen2022,nyhegn2022,nyhegn2023}, utilizing a generalized Holstein-Primakoff. This relies on the system supporting AFM order, such that we can define two sublattices where the spins predominantly point up on sublattice A and down on sublattice B. Using this, we express the operators on sublattice A as $\hat{ S }^{z}_{l,{\mathbf i}} = (1-\hat{ h }^{\dagger}_{l,{\mathbf i}}\hat{ h }_{l,{\mathbf i}})/2 - \hat{ s }^{\dagger}_{l,{\mathbf i}}\hat{ s }_{l,{\mathbf i}} $ and $\hat S_{l,{\mathbf i}}^{-}= \hat S^x_{l,{\mathbf i}}-i\hat S^y_{l,{\mathbf i}}=\hat{ s }^{\dagger}_{l,{\mathbf i}} (1-\hat{ s }^{\dagger}_{l,{\mathbf i}}\hat{ s }_{l,{\mathbf i}}-\hat{ h }^{\dagger}_{l,{\mathbf i}}\hat{ h }_{l,{\mathbf i}})^{1/2}$, and the creation operators are $\tilde{c}_{l,\bi,\Dn} = \hat{ h }^{\dagger}_{l,{\mathbf i}}\hat{ s }_{l,{\mathbf i}}$ and $\tilde{c}_{l,\bi,\Up} = \hat{ h }^{\dagger}_{l,{\mathbf i}}(1-\hat{ s }^{\dagger}_{l,{\mathbf i}}\hat{ s }_{l,{\mathbf i}}-\hat{ h }^{\dagger}_{l,{\mathbf i}}\hat{ h }_{l,{\mathbf i}})^{1/2} $. Opposite to the slave-boson description used in the main text, $\hat{ h }^{\dagger}_{l,{\mathbf i}}$ is a fermionic creation operator of a hole, and $\hat{ s }^{\dagger}_{l,{\mathbf i}}$ is a bosonic creation operator describing derivation from the AFM ordered state. Similar relations are used for sublattice B.

\begin{figure}[t!]
	\begin{center}
	\hspace{-0.5cm}
	\includegraphics[width=0.4\textwidth]{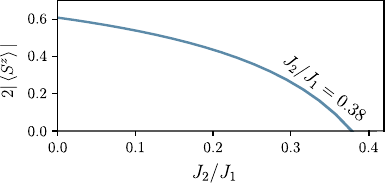}
	\end{center}
	\caption{The average magnetization, $2| \left\langle S_{z}\right\rangle | = 1 - 2 \sum_{\bk}v^{2}_{\bk}/N $, calculated within LSWT. This predicts a phase transition to occur at $J_{2}/J_{1} = 0.38$.}
	\label{Fig.sz}
\end{figure}

Using these expressions and keeping only linear terms, we can diagonalize the $J$ part of the Hamiltonian to find
\begin{align}
	\hat{ H  }_J = \sum_{\bk}\omega^{m}_{\bk} \hat{ b}^{\dagger}_{\bk}\hat{b}_{\bk} + E_{0},
\end{align}
with
\begin{align}
	\omega^{m}_{\bk} &= \ \frac{1}{2} \sqrt{\tilde{J}^{2}- (\alpha_{1}J_{1}z\gamma_{\bk} )^{2}}, \nonumber \\
	\tilde{J}_{\bk} &\equiv z\left( J_{1} + J_{2}\left[ \alpha_{2} \gamma'_{\bk} - 1 \right]\right),
\label{Eq:dispersion_rel}
\end{align}
for $z=4$ and $\gamma_{\bk} = \sum_{\boldsymbol{\delta}}e^{i\bk\cdot \boldsymbol{\delta}}/z$ and $\gamma'_{\bk} = \sum_{\boldsymbol{\delta}'}e^{i\bk\cdot \boldsymbol{\delta}'}/z$ being the structure factors for NN and NNN terms, respectively. The spin wave operators, $\hat{b}$, are related to the spin fluctuations, $\hat{s}$, by a Bogoliubov transformation
\begin{align}
	\begin{bmatrix}
   		\hat{s}_{\bk} \\ 
		\hat{s}^{\dagger}_{-\bk} 
	\end{bmatrix} = \begin{bmatrix}
   		u_{\bk} & -v_{\bk} \\
    		-v_{\bk} & u_{\bk}
	\end{bmatrix}
	\begin{bmatrix}
   		\hat{b}_{\bk} \\ 
		\hat{b}^{\dagger}_{-\bk} 
	\end{bmatrix},
	\label{Eq:mapping}
\end{align}
where $u_{\bk} = \sqrt{\frac{1}{2}\left(\tilde{J}/(2\omega^{m}_{\bk}) + 1\right)}$ and $v_{\bk} =  {\rm sgn}\left[\gamma_\bk \right]\sqrt{\frac{1}{2}\left(\tilde{J}/(2\omega^{m}_{\bk}) - 1\right)}$ are the coherence factors. Looking at Fig. \ref{Fig.sz}, we see that this approach qualitatively captures the behavior when $J_{2}$ is increased. This predicts a phase transition to occur at $J_{2}/J_{1}=0.38$, which is not far from the predictions made with more advanced methods. This yields it to being around $J_{2}/J_{1} \sim 0.46$ \cite{liu2022, wang2018, nomura2021, ferrari2020}.

\begin{figure}[t!]
	\begin{center}
	\includegraphics[width=0.48\textwidth]{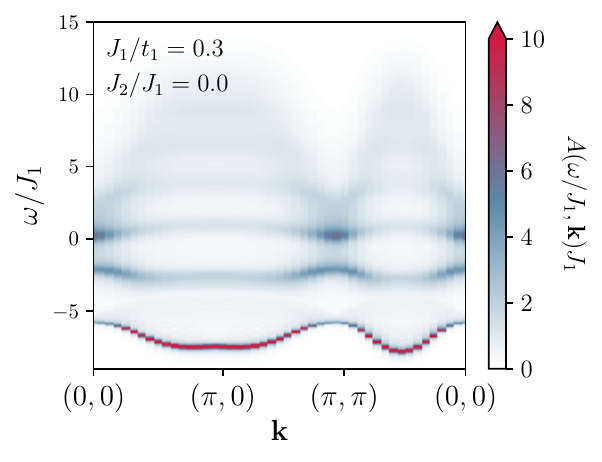}
	\end{center}
	\caption{Spectral function of a hole along the reversed path in the Brillouin zone as in Fig.~\ref{Fig.spec1}
 for $J_{1}/t_{1}=0.3$, $J_{2}/J_{1}=0$.}
	\label{fig.BZcompare}
\end{figure}

With the described transformations, the kinetic term takes the form
\begin{align}
	\hat{H}_t= &\sum_{\mathbf{k}} \omega^{p}_{\bk}  \hat{h}_{\mathbf{k}}^{\dagger}\hat{h}_{\mathbf{k}} + \sum_{\bk,\bp}  g(\bp,\bk)\hat{ h }_{\bp+\bk}^{\dagger}\hat{ h }_{\bp} \left( \hat{ b}^{\dagger}_{-\bk} +  \hat{b}_{\bk} \right) 
	\label{Eq:HtSlaveFermion}
\end{align}
with 
\begin{align}
	g(\bp,\bk) = \frac{zt_{1}}{\sqrt{N}}\left[u_{\bk} \gamma_{\bp + \bk} - v_{\bk}\gamma_\bp\right],
	\label{Eq:g_vertices}
\end{align}
and 
\begin{align}
	\omega^{p}_{\bk} =2t_{2}\left[ \cos{(k_{x} + k_{y})} + \cos{(k_{x} - k_{y})}\right].
	\label{Eq:disp}
\end{align}
For a finite $J_{2}$ we hence get a bare dispersion corresponding to the hole jumping within the same sublattice. Proceeding as done in the literature \cite{kane1989,martinez1991,liu1991,schmitt-rink1988,nielsen2021,nielsen2022,nyhegn2022,nyhegn2023}, we can use the above expressions to calculate the Green's function, 
\begin{align}
	G(\bk,\tau) = -i\left\langle T_{\tau} \left[ \hat{h}_{\bk}(\tau) \hat{h}^{\dagger}_{\bk}(0)\right]\right\rangle,
	\label{Eq:Greens_SCBA}
\end{align}
of a single hole with the SCBA approximation. The spectral function of such result is presented in Fig. \ref{fig.BZcompare} for $J_{2}/J_{1}=0$ and $J_{1}/t_{1}=0.3$, which shows good agreement with the MPS result presented in Ref. \cite{bohrdt2020b}.

\begin{figure}[t!]
	\begin{center}
	\includegraphics[width=0.48\textwidth]{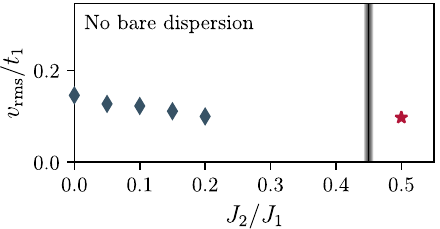}
	\end{center}
	\caption{Same plot as in Fig. \ref{Fig.1}, but where we have omitted the bare dispersion of the hole in the AFM phase. As we approach the phase transition from the AFM side, the hole experiences a slowdown instead of a speedup as in Fig. \ref{Fig.1}.}
	\label{fig.NoDisp}
\end{figure}

Investigating the nature of the rms speedup as we approach the phase transition, we find that by neglecting the bare dispersion of the hole we retrieve an rms velocity which decreases with interaction strength; see Fig. \ref{fig.NoDisp}. This we ascribe to the fact that with an increasing $J_{2}/J_{1}$ the spinon spectrum gets increasingly softer around $\bp = (\pi,0)$ and $\bp = (0,\pi)$.

\section{String excitations for hole}  \label{app.hole.string}

\begin{figure}[t!]
	\begin{center}
	\includegraphics[width=0.48\textwidth]{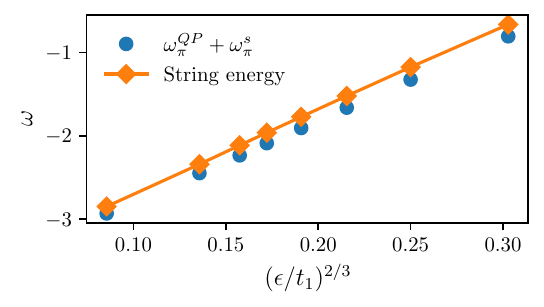}
	\end{center}
	\caption{Excitation energy of the excitation peak from the hole spectral function for $\bp=(0,0)$ in orange. Equivalent to Eq. \eqref{Holedispersion}, the blue data points show the energy of the holon excitation peak for $\bp=(\pi,\pi)$ added to $\omega_{(\pi,\pi)}$ coming from the Van Hove singularity discussed in the main part. This is plotted as a function of $(\epsilon/t_{1})^{2/3}$.}
	\label{fig.string.hole}
\end{figure}

In this appendix, we show that indeed the higher excitation peaks in the hole's spectral function have the characteristic two-thirds scaling associated with string excitations. This is seen in Fig. \ref{fig.string.hole}, where we plot the energy of the peak as a function of $(\epsilon/t_{1})^{2/3}$. Here, we see the linear behavior and in addition, we see that the energy is given by the energy of the holon string peak plus $\omega_{(\pi,\pi)}$ corresponding to the Van Hove singularity discussed in the main part. This shows that the excitation peaks in the hole spectral function come from the string excitations of the holon.

\section{Quasiparticle band fit}  \label{app.QP.fit}

Interestingly, we find that the lowest quasiparticle band can be fitted to the form 
\begin{align}
	\omega^{\rm QP}_{\bk} =&\tilde{t}_{1} (\cos{k_{x}} + \cos{k_{y}} ) 
	+\tilde{t}_{2} [ \cos{(k_{x}+k_{y})} + \cos{(k_{x}-k_{y})} ] \nonumber \\
	&+  \tilde{t}_{3} ( \cos{2k_{x}} + \cos{2k_{y}} ),
\end{align}
shown as the red dashed line in Fig. \ref{Fig.spec2}(b), where $\tilde{t}_{2}/\tilde{t}_{1} \approx 0.15$ and $\tilde{t}_{3}/\tilde{t}_{1} \approx -0.05$. We find that these ratios are approximately constant for strong interaction strengths $\epsilon/t_{1}<0.1$.

\clearpage

\bibliography{SpinLiquid}

\end{document}